\documentstyle[mncite,draft,psfig]{mn}

\title[J-band spectra of CVs]{J-band spectroscopy of
 cataclysmic variables}

\author[S.\,P.\ Littlefair et.\ al.]{S.\,P.\ Littlefair,$^{1}$
 V.\,S.\ Dhillon,$^{1}$ Steve B. Howell$^{2}$ and David R. Ciardi$^3$\\
$^1$Department of Physics and Astronomy, University of Sheffield, 
Sheffield S3 7RH, UK \\
$^2$Department of Physics and Astronomy, University of Wyoming, 
Laramie WY 82071, USA\\
$^3$Department of Astronomy, University of Florida, Gainesville, FL 32611, 
USA\\}

\date{\center{\Large Submitted for publication in the Monthly Notices of the
Royal Astronomical Society \\ 
\vspace{.5cm} \today}} 

\begin{document}
\maketitle

\begin{abstract} 
We present time resolved J-band 
(1.025 -- 1.340 $\mu$m) infrared 
spectra of the short-period dwarf novae (DNe) WZ~Sge, VY~Aqr and single
spectra of the short-period DNe EF~Peg and
the novalike variable PX~And. There is some evidence in the spectra of 
VY~Aqr and EF~Peg that we have detected the
secondary star, both in the continuum slope and also through
the possible presence of spectral features. 
The spectra of WZ~Sge and PX~And, on the other hand, show no evidence for
the secondary star, with upper limits for it's contribution
to the J-band light of 10\% and 20\% respectively.
The spectral type of the secondary in WZ~Sge is constrained to be 
later than M7.5V.
Using skew mapping we have been able to derive a value for the radial
velocity semi-amplitude of the secondary star in VY~Aqr of $K_R$ 
= $320\pm70$ kms$^{-1}$, which in
conjunction with $K_W$ from \scite{thorstensen97} gives a mass ratio of
$q=0.15\pm0.04$.
\end{abstract} 


\section{Introduction}
\label{sec:introduction}

Cataclysmic variables (CVs) are semi-detached binary systems in which a white
dwarf primary accretes material from a Roche-lobe filling secondary. For a 
thorough review of CVs see \scite{warner95a}. 
There is a strong motivation to detect the secondary stars in CVs. A detection
of the secondary star is desirable for a reliable estimation of the 
system parameters, which in turn allow one to
test theories of CV evolution, (e.\ g.\ see \pcite{sad98}). In addition the 
contribution of the secondary to the total flux can be used to give the 
distance to the CV (e.\ g.\ see \pcite{marsh90c}).
Detection of the secondary star is of particular interest in CVs
below the period gap. The models of \scite{kolb93} predict that 99\% of the 
present-day population of CVs should be below the period gap and that around 
70\% of these systems will have already reached the orbital period minimum. 
Modelling of the secondary stars in post-period-minimum CVs has been 
performed by \scite{howelletal97}, who find them to be 
degenerate, brown dwarf-like objects with masses 
between 0.02-0.06 $M_\odot$.
\scite{howelletal97} further speculate that the tremendous 
outburst amplitude dwarf novae (TOADs) are these post-period-minimum CVs.
Estimates of the secondary stars' parameters would provide us with
tests for these claims, as well as allowing us to investigate the mass-radius
relation of the brown dwarf-like secondary by invoking the Roche lobe-filling
criterion.

Previous attempts to detect the secondary star in CVs have focussed on the TiO
bands and Na{\small I} lines in the optical-infrared (0.7 -- 1.0 
$\mu$m), e.g \scite{smith88}. 
These searches proved insensitive to the secondary star
in short-period CVs, although there were exceptions (e.g Z Cha,
\pcite{wade88}). The relative insensitivity of optical searches to the 
secondary in 
short-period CVs is unsurprising, as short period CVs are expected to possess 
late-type M-dwarf secondary stars\footnote{The spectral type of the secondary 
star in a CV can be estimated from it's orbital period, $P$, using the 
relationships $26.5-0.7P$ (for $P<4$ hr) and $33.2-2.5P$ (for $P>4$ hr), 
where $G0=0$, $K0=10$ and $M0=20$ \cite{sad98}}. 
These late-type M-dwarf stars have spectra which 
peak in the infrared (see \pcite{kleinmann86}), and at optical
wavelengths the light from the M-dwarf is swamped by light from the 
accretion disc.
Infrared searches for the secondary star in short period CVs have largely 
focussed on the K-band \cite{dhillon97a}, with some success. 
Whilst the secondary star has proved relatively simple to
find in dwarf novae above the period gap, where the contribution of the 
secondary in the K-band is typically $>75\%$ \cite{dhillon97a}, it has 
proved much more difficult to detect the secondary star in both nova-likes
and dwarf novae below the period gap. For these systems the contribution
of the secondary to the infrared flux is usually much less (typically 
less than 25\%; \pcite{dhillon97a}). 
To find the secondary stars in these systems a new strategy is required. In 
this paper we present the first results of a program to detect secondary star 
features using the J-band spectra of cataclysmic variables.

Searching for secondary star features in the J-band has certain advantages
over other wavelength ranges. First, observations in the near-infrared are 
plagued by a very high background. This is due to both thermal emission
from the telescope, sky  and instrument, and atmospheric emission from 
molecules (mostly $OH^-$ and $O_2$) excited by daytime solar radiation, 
resulting in airglow which is both temporally and spatially variable 
\cite{ramsay92}. This background is significantly lower in the J-band 
than in K (16 mag/arcsec in J compared with 12 mag/arcsec in K - data for 
Mauna Kea, Hawaii). 
Second, in the late-type M-dwarfs expected in post-period gap CVs, there are 
more absorption features in the J-band than K, and these absorption features 
have larger equivalent widths than those in the K-band (e.g see 
\pcite{jones94}).
Third, the change in strength of absorption features with  
spectral type in late-type M-dwarfs is more marked in the J-band than in
K (c.f.\ table 2 in \pcite{dhillon95a} with table \ref{tab:lines} in this 
paper). This high sensitivity of absorption features to 
spectral type makes a secondary star detection in the J-band a powerful 
constraint on models of the evolution of CVs below the period gap.

\section{Observations}
\label{sec:observations} 

On the nights of 1998 August 8 \&\ 9 we obtained spectra of the Dwarf Novae 
(DNe) WZ~Sge, EF~Peg, VY~Aqr, the novalike variable PX~And and the 
M-dwarfs Gl644C and Gl866AB
with the Cooled Grating Spectrometer (CGS4) on the 3.8~m
United Kingdom Infrared Telescope (UKIRT) on Mauna Kea, Hawaii. CGS4 is 
a 1--5 micron spectrometer containing an InSb array with 256$\times$256 pixels.
The 40~l/mm grating with the 300~mm camera gave a resolution of 
approximately 300~km\,s$^{-1}$. To cover the wavelength range 1.025--1.340 
microns required one grating setting, centred at 1.175 microns (second order).
Optimum spectral sampling and bad pixel removal were obtained by mechanically 
shifting the array over two pixels in the dispersion direction in 
steps of 0.5 pixels.  
We employed the non-destructive readout mode of the detector to reduce
the readout noise. In order to compensate for fluctuating atmospheric 
OH$^-$ emission lines we took relatively short exposures (typically 60s) 
and nodded the telescope primary so that the object spectrum switched between 
two different spatial positions on the detector.
The slit width was 0.6 arcseconds (projecting to approximately 1 pixel 
on the detector) and was oriented at the parallactic 
angle throughout the second night. Observations of WZ~Sge and Gl866AB were
taken with a slit position angle of $0^\circ$ to avoid light from a star 
close to WZ~Sge falling on the slit. Using the UKIRT tip-tilt secondary, the seeing was around 
$0.5''$ on the first
night, and dropped steadily from $0.8''$ to $0.4''$ on the second night.
The humidity was low (10-20\%) throughout, and the sky was photometric for 
most of the run, although some high cirrus developed in the second half of 
the  first night. A full journal of observations is presented  in 
table~\ref{tab:journal}. 

\section{Data Reduction} 
\label{sec:datared} 

The initial steps in the reduction of the 2D frames were performed
automatically by the CGS4 data reduction system \cite{daley94}. 
These were: the application of
the bad pixel mask, bias and dark frame subtraction, flat field division,
interlacing integrations taken at different detector positions, and co-adding
and subtracting nodded frames. Further details of the above procedures may be
found in the review by \scite{joyce92}. In order to obtain
1D data, we removed the residual sky by subtracting a polynomial fit and
then extracted the spectra using an optimal extraction technique 
\cite{horne86a}.
The next step was the removal of the ripple arising
from variations in the star brightness between integrations (i.e. at different
detector positions). These variations were due to changes in the seeing, sky
transparency and the slight motion of the stellar image relative to the slit.

There were two stages to the calibration of the spectra. The first was the
calibration of the wavelength scale using krypton arc lamp exposures. The
second-order polynomial fits to the arc lines  yielded an error of less
than 0.0001 microns (rms).  The final
step in the spectral calibration was the removal of telluric 
features and flux calibration. This was performed by dividing the spectra to 
be calibrated by the spectrum of an F-type standard, with its prominent 
stellar features interpolated across. F-types were taken throughout the 
night, at different airmasses. In each case, the F-star used was that which 
gave the best removal of telluric features, judged by the cancellation of the 
strong features around 1.14 microns.
We then multiplied the result by the known flux of the standard at each
wavelength, determined using a black body function set to the same 
effective temperature and flux as the standard. As well as providing flux
calibrated spectra, this procedure also removed telluric absorption features
from the object spectra. 

\section{Results}
\label{sec:results}

\begin{figure*}
\centerline{\psfig{figure=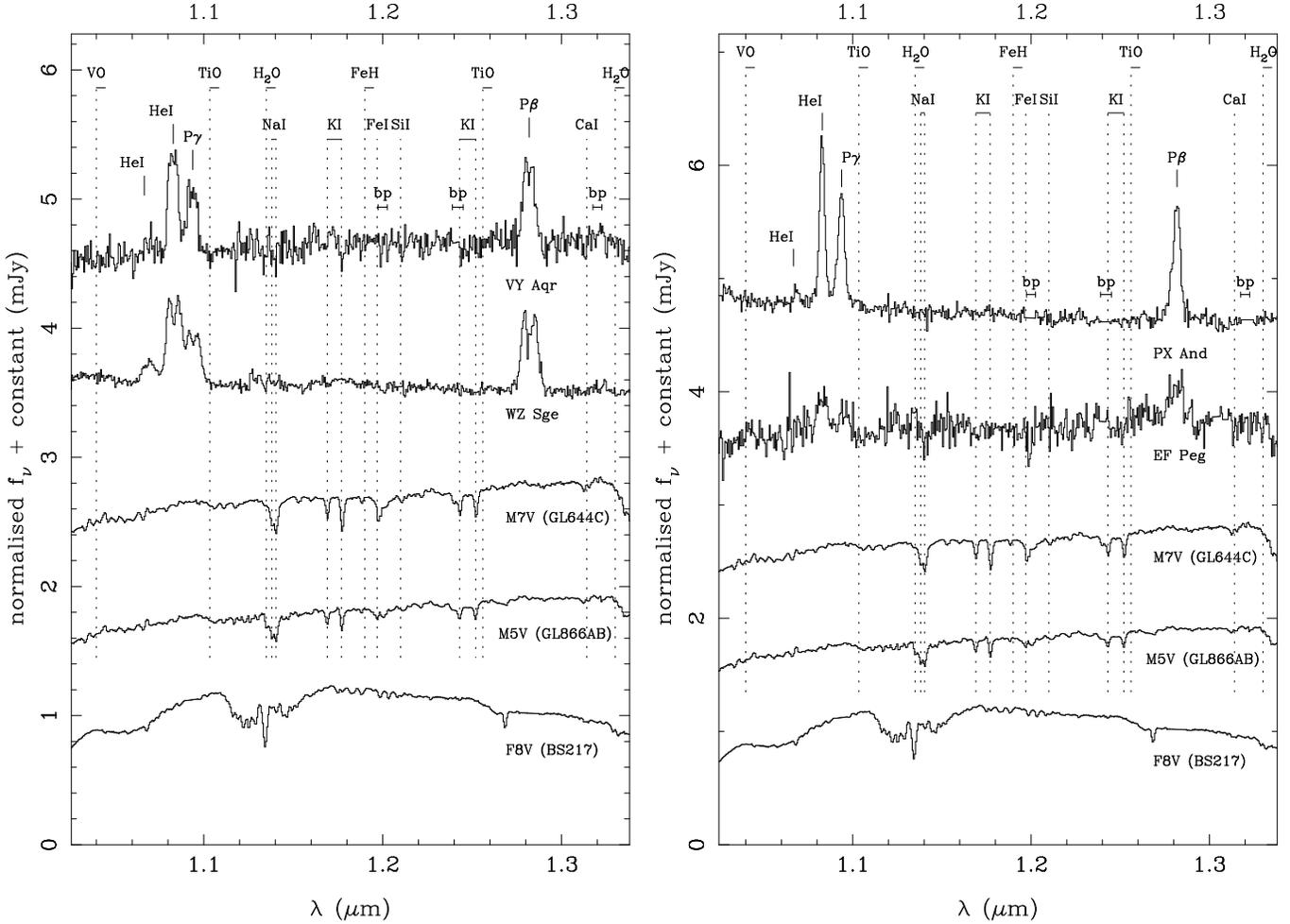,width=18.0cm}}
\caption{Left-hand side: Average J-band spectra of the short period DNe 
WZ~Sge, VY~Aqr and the M-dwarfs Gl644C and Gl866AB.
Right-hand side: Average J-band spectra of the short period DN
EF~Peg, the novalike variable PX~And and the M-dwarfs Gl644C and Gl866AB.
For both plots, the spectra have been normalized by dividing by the flux 
at 1.3 $\mu$m and then offset by adding a multiple of 0.9
to each spectrum. Also shown is the spectrum of an F8V star,
normalized by dividing by the flux at 1.3 $\mu$m, which
indicates the location of telluric features.}
\label{fig:cvav}
\end{figure*}

\begin{figure}
\psfig{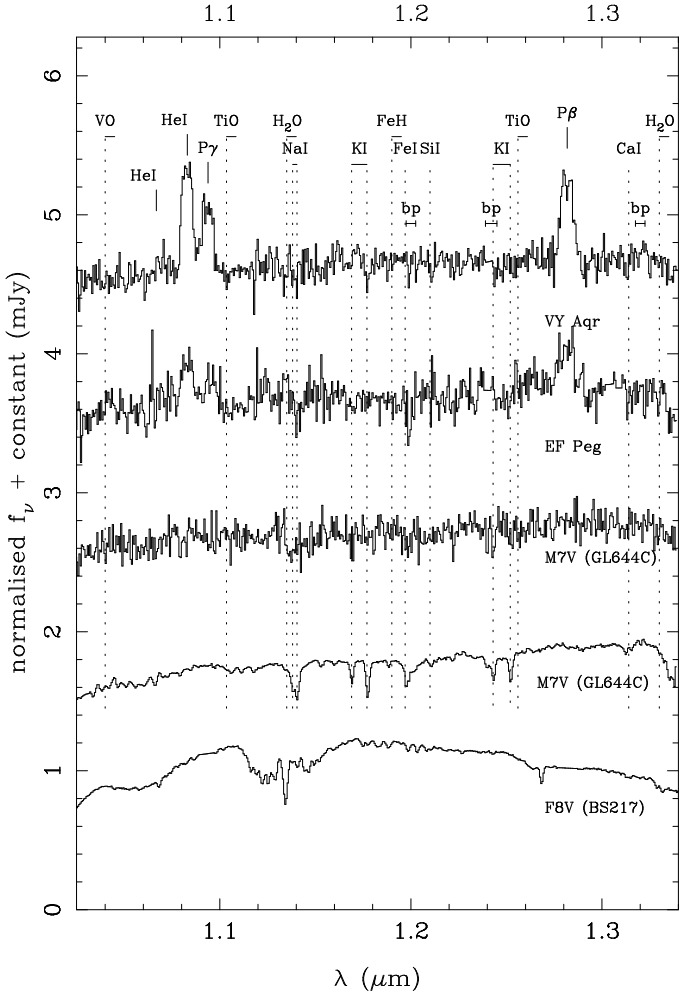}
\caption{Average J-band spectra of the short period DNe VY~Aqr, EF~Peg  
and the M-dwarf Gl644C. 
The spectra have been normalized by dividing by the flux 
at 1.3 $\mu$m and then offset by adding a multiple of 0.9
to each spectrum. One spectrum of Gl644C has been divided by 1.7, 
in an attempt to match the appearance of the absorption feature at 1.33 $\mu$m
 to that seen in VY~Aqr and noise has then been added to the spectrum in order
to match the signal-to-noise to that of the DNe spectrum. 
Also shown is the spectrum of an F8V star, normalized by dividing by the 
flux at 1.3 $\mu$m, which indicates the location of telluric features.}
\label{fig:noisy}
\end{figure}

Figure~\ref{fig:cvav} shows the 1.025--1.340~$\mu$m averaged spectra
of the short period DNe WZ~Sge and VY~Aqr (left hand side) and the 
1.025--1.340~$\mu$m averaged spectra
of the short period DN EF~Peg and the novalike variable PX~And (right hand 
side) together with the spectra of M7 and M5 field dwarfs.
We also show the spectrum of an F8V star, which 
indicates the location of telluric absorption features; spectral features 
within the strongest absorption bands are highly uncertain. 
The location of bad pixels are labelled by a bar showing the extent of the
anomaly and the label `bp'.
In table~\ref{tab:lines} we list the 
wavelengths, equivalent widths and velocity widths of the most 
prominent spectral lines identified in figure~\ref{fig:cvav}.

\subsection{Accretion disc features}
\label{subsec:disc}
The spectra of WZ~Sge and VY~Aqr (fig. \ref{fig:cvav}) are dominated by the 
strong, double-peaked 
emission lines of He{\small I}, Paschen-$\gamma$ and Paschen-$\beta$. 
Also seen is a small broad peak around 1.07 $\mu$m. This is identified as the 
He{\small I} line at 1.0668 $\mu$m.
The broad, double-peaked nature of these lines indicates an origin in the 
accretion disc.
The full-width at half-maxima (FWHM) for WZ~Sge are in agreement with those
presented by \scite{dhillon99}, who find a FWHM for the
Brackett-$\gamma$ line of 2200 kms$^{-1}$. FWHM for VY~Aqr are in good 
agreement with \scite{thorstensen97}, who find a FWHM of 1550 kms$^{-1}$
for H$\alpha$. 
We measured the separation of the double-peaked emission lines for 
VY~Aqr and WZ~Sge. For WZ~Sge the peak separations were 
$1400\pm 100$ kms$^{-1}$, $1300\pm 100$ kms$^{-1}$  and $
1400\pm 100$ kms$^{-1}$ for He{\small I}, P$\gamma$ and P$\beta$, 
respectively. These values are in good agreement with values for the
peak separation derived from optical data, e.g.\ 1380 kms$^{-1}$ for H$\alpha$
\cite{gilliland86}.
For VY~Aqr the peak separations were $700\pm 100$ kms$^{-1}$, 
$800\pm 150$ kms$^{-1}$ and $900\pm 100$ kms$^{-1}$ 
for He{\small I}, P$\gamma$ and P$\beta$, respectively. These peak separations 
for VY~Aqr are in good agreement with \scite{thorstensen97}, who find a peak 
separation of 930 kms$^{-1}$ for H$\alpha$.

The spectrum of PX~And (fig. \ref{fig:cvav}) is dominated by the strong, 
single-peaked emission lines
of He{\small I}, Paschen-$\gamma$ and Paschen-$\beta$, typical of the SW~Sex
stars of which PX~And is a member \cite{still95b}. The FWHM for PX~And are 
consistent with the values given by
\scite{still95b}, who find FWHM of 930 -- 1030 kms$^{-1}$ for
He{\small I} emission and 1080 -- 1600 kms$^{-1}$ for the Balmer lines. 

The spectrum of EF~Peg (fig. \ref{fig:cvav}) also shows strong emission 
lines, although the
signal-to-noise ratio is too low to say if these are single or double-peaked.
The large velocity widths of the emission lines in EF~Peg, suggests an 
accretion disc origin.

\subsection{Secondary star features}
\label{subsec:secondary}

The spectrum of VY~Aqr shows increasing flux towards the red end of the 
spectrum, with a continuum slope of $(7.4\pm0.4) \times10^{-1}$ mJy/$\mu$m, 
suggesting
that the secondary star makes a significant contribution to the continuum 
light in VY~Aqr (c.f. the slope of the M-dwarfs in figure \ref{fig:cvav}).
The continuum in VY~Aqr also exhibits a change in slope at $\sim$ 1.1$\mu$m.
This feature is also present in the spectra of the M-dwarfs.
The spectrum of VY~Aqr also shows a tentative detection of secondary star 
absorption features, particularly the headless water band seen at 1.33 $\mu$m.
Likewise, the EF~Peg continuum shows increased flux at the red end of the 
spectrum, with a continuum slope of $(5.7\pm0.8) \times10^{-1}$ mJy/$\mu$m and
like the spectrum of VY~Aqr, EF~Peg also shows a
tentative secondary star detection through the water band at 1.33 $\mu$m.
The EF~Peg continuum also shows a change in slope at $\sim$ 1.1$\mu$m, in
common with VY~Aqr and the M-dwarfs.
The detection of the water band in EF~Peg and VY~Aqr should be considered 
carefully: the spectra do
not show any of the other absorption features visible in the spectra of the 
M-dwarfs -- although it is possible they are lost in the noise.
Figure \ref{fig:noisy} shows the spectra of the DNe EF~Peg and VY~Aqr, 
together with two spectra of the M-dwarf Gl 644c. One spectrum of Gl 644c
has been divided by 1.7 in order to weaken the appearance of the spectral
lines (the correct amount was judged by attempting to match the appearance
of the 1.33$\mu$m water feature to that seen in VY~Aqr). This spectrum has 
then had noise added in order to match the signal-to-noise
to that of the DNe spectra. Absorption features are barely discernable in this
spectrum. This shows that absorption features from the secondary star are
probably hidden by noise in the CV spectra.
Furthermore, the most prominent 
absorption features fall at the same wavelengths as the strongest telluric 
features and bad pixels, and any features in these regions are inevitably 
more uncertain. 
In addition, the water feature at 1.33 $\mu$m only occurs in concurrence with
the red continuum seen in VY~Aqr and EF~Peg, and does not appear in the 
spectrum of WZ~Sge, for example. We believe that this constitutes a body of 
evidence for the detection of the secondary star in VY~Aqr and EF~Peg. This is
discussed further in section 5.1. Unfortunately, without detection of more 
absorption features and the observation of more field-dwarf templates we are 
unable to determine the spectral type of the secondary star in EF~Peg and 
VY~Aqr.

There is no evidence for the secondary star in the spectra of WZ~Sge and 
PX~And, which show a blue continuum without any changes in slope or 
absorption features.

\subsection{Secondary star contributions and distances}
\label{subsec:dist} 

Table \ref{tab:contrib} shows the contribution of the secondary star to the
total flux of each CV. Percentages
were calculated by normalising the average spectra of both the CV and M-dwarf 
template. An increasing fraction of the M-dwarf spectrum was then subtracted 
from the CV spectrum until spectral features in the CV spectrum were removed
or absorption features from the M-dwarf appeared in 
emission. The fraction at which this occurs gives the 
contribution of the secondary star to
the total flux. Where no secondary star features are present this represents
an upper limit to the contribution of the secondary star.

The contributions were found to vary depending upon
which M-dwarf template was used. It follows that systematic error could be
introduced through using a template whose spectral type differs greatly from 
the actual spectral type of the secondary star.

The strongest features in the template spectra increase in strength with
spectral type. Hence templates of earlier spectral type give a higher 
contribution than those later in the spectral sequence. Hence, for WZ~Sge and 
PX~And, where the earliest spectral type consistent with the spectral 
type-orbital period relation in \scite{sad98} was used, we have derived 
strict upper limits to the contribution of the secondary star to the J-band
light.

In VY~Aqr and EF~Peg, we used the M-dwarf which was able to correct the H$_2$0
band at 1.33 $\mu$m and flatten the continuum without introducing spectral 
features in emission. It would have been apparent if the spectral type of
the template used was too late, as the KI feature at $\sim$ 1.17 $\mu$m 
increases in strength more rapidly with spectral type than the H$_2$0 feature
at 1.33 $\mu$m. Hence a template with too late a spectral type would not be
able to correct the water feature without introducing the KI feature in 
emission. Templates whose spectral types are too early would give a larger
value for the secondary star contribution. However, the spectral type of the 
secondary in EF~Peg and VY~Aqr should be later than M5V. 
Hence we believe the contributions derived for these stars are reliable. 
The errors shown in table \ref{tab:contrib} are the discrepancies 
between the values obtained with the two M-dwarf templates and hence reflect
the uncertainty resulting from the unknown spectral type.

The contributions of the secondary star will also be affected by systematic
error due to flux from the disc if an emission feature corrupts the water 
band at 1.33 $\mu$m. This is not thought to be the case. 
 
We find an upper limit to the secondary star contribution to the J-band light 
in WZ~Sge of $10\%$, consistent with \scite{ciardi98}, who model the 
relative  contributions to the near-infrared emission in WZ~Sge and find that 
the secondary star should contribute $\approx 20\%$ of the near-infrared 
flux, although their models suggest that this figure is lower 
(perhaps $\approx 10\%$) in the J-band.

If the J-band magnitude for the CV is known it is possible to
derive the apparent J-band magnitude of the secondary. In principle,
this apparent magnitude can then be used to estimate the distance to the CV,
by comparison with the absolute magnitude of a field-dwarf of appropriate
spectral type.
In the cases where the secondary has not been detected, this
value represents a lower limit to the distance involved. 
J-band magnitudes for WZ~Sge, VY~Aqr and PX~And are 14.2, 15.3 and 14.2 
respectively (Mark Huber, private correspondence). The J-band magnitude of
EF~Peg is 16.1 \cite{sproats96}. 
This gives a J-band magnitude of $15.7\pm0.5$ for
the secondary in VY~Aqr, $16.7\pm0.5$ for the secondary in EF~Peg and upper 
limits of 16.2 and 15.4 for the J-band magnitudes of the secondaries in 
WZ~Sge and PX~And respectively. 
Unfortunately, we are unable to give distances to these CVs, as the absolute
J-band magnitude of M-dwarfs varies rapidly with spectral type -- c.f.\ 
M$_J = 6.56$ for M5V with M$_J = 8.75$ for M7V   
\cite{bessel91} -  a reliable spectral type determination is necessary 
before distances can be estimated.
As an aside, it is worth noting that both \scite{spruit98} and \scite{smak93}
find a distance to WZ~Sge of 48 pc, which, through the analysis above, and
the data in \scite{bessel91} would constrain the spectral type of the secondary
to be later than M7.5V, consistent with \scite{ciardi98}, who find
that the secondary star in WZ~Sge is cooler than 1700K.

\subsection{Time resolved spectra}
\label{subsec:timeres}

\begin{figure*}
\centerline{\psfig{figure=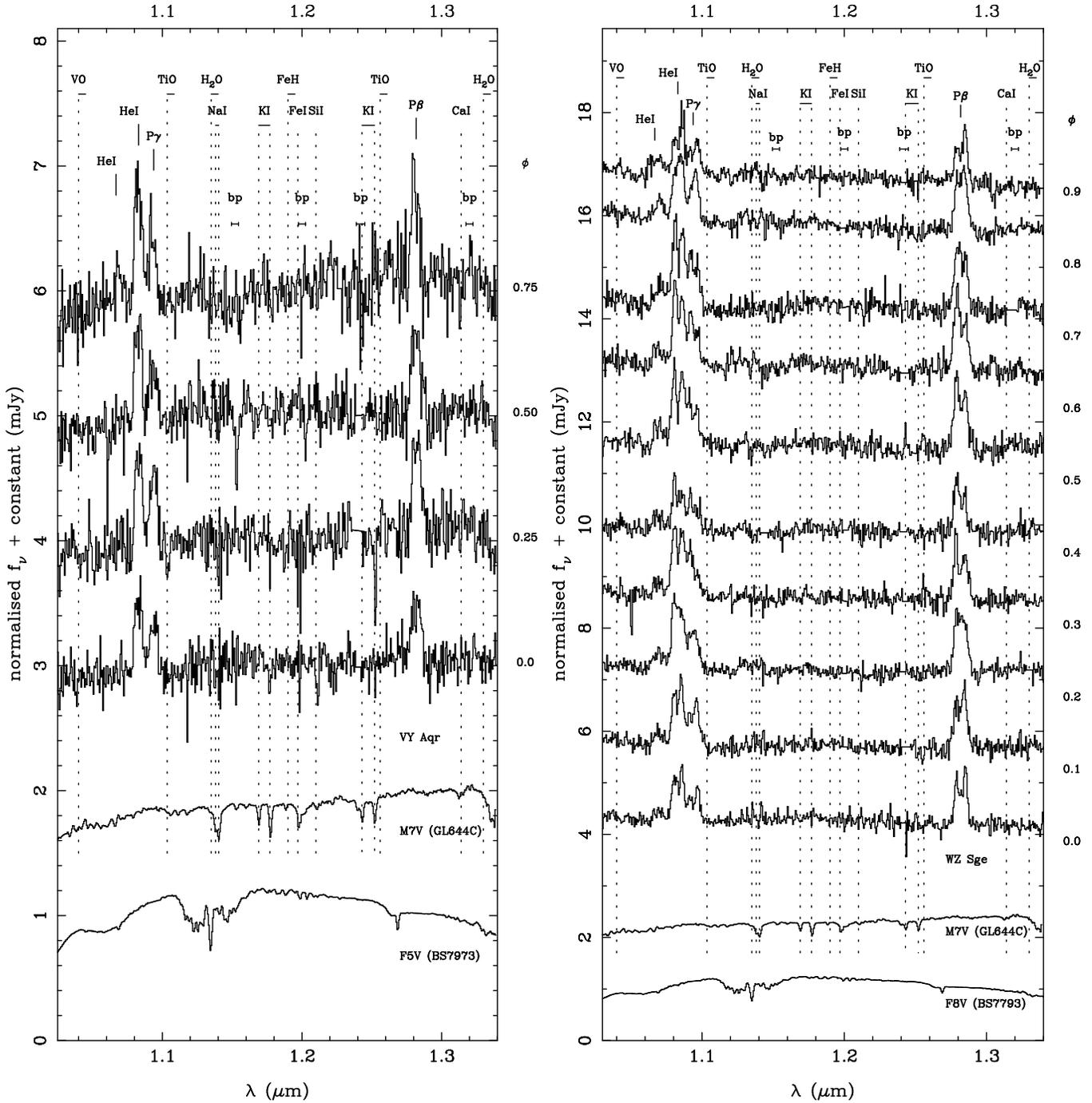,width=18.0cm}}
\caption{Left-hand side: Time-resolved J-band spectra of the short period 
DN VY~Aqr and spectra of the M-dwarf Gl644C. 
Orbital phase is shown on the right-hand side of the plot, using the
ephemeris of \protect\scite{thorstensen97} -- i.e without applying the phase
offset described in section 4.5.
Right-hand side: Time-resolved J-band spectra of the short period 
DN WZ~Sge and spectra of the M-dwarf. 
Orbital phase is shown on the right-hand side of the plot, using the
ephemeris of \protect\scite{spruit98}.
In both plots the spectra have been normalized by dividing by the flux 
at 1.3 $\mu$m. The spectra were then offset by a multiple of 1.0 for VY~Aqr
and 1.4 for WZ~Sge.
Also shown is the spectrum of an F8V star, normalized by dividing by the 
flux at 1.3 $\mu$m, which indicates the location of telluric features. 
The location of bad pixels is labelled by a bar showing the extent of the
anomaly and the label `bp'.
}
\label{fig:tr}
\end{figure*}

Figure~\ref{fig:tr} shows the J-band time-resolved 
spectra of VY~Aqr (left-hand side) and  the J-band time resolved spectra of 
WZ~Sge (right-hand side) together with the spectra of M7 and M5 field dwarfs. 
The majority of ephemerides for WZ~Sge are for mid-eclipse of what is 
believed to be the bright spot. In this paper, WZ~Sge is phased according to 
the ephemeris of \scite{spruit98}, whose binary phase zero is corrected to 
represent mid-eclipse of the white dwarf.

The emission lines in VY~Aqr vary considerably over the orbit. 
The H$_2$0 feature at 1.33$\mu$m also shows variation with phase. It does not
appear to be present at phase $\phi=0.0$. We have been
unable to account for this variation; possible causes are discussed in section
5.1.
There is no evidence for a change in the continuum slope in VY~Aqr with 
orbital phase, with the slope always remaining within one standard error
of the mean, given in section 4.2.

The emission lines in WZ~Sge
show a strong variation throughout the orbit. There is a reduction in line
strength at $\phi \approx 0.9$ and $\phi \approx 0.0$, corresponding to
primary eclipse. The lines appear single peaked at $\phi \approx 0.3$ and
$\phi \approx 0.8$.

Time resolved spectra were not obtained for EF~Peg and PX~And.

\subsection{Skew mapping and Doppler tomography}
\label{subec:mapping}

\begin{figure*}
\centerline{\psfig{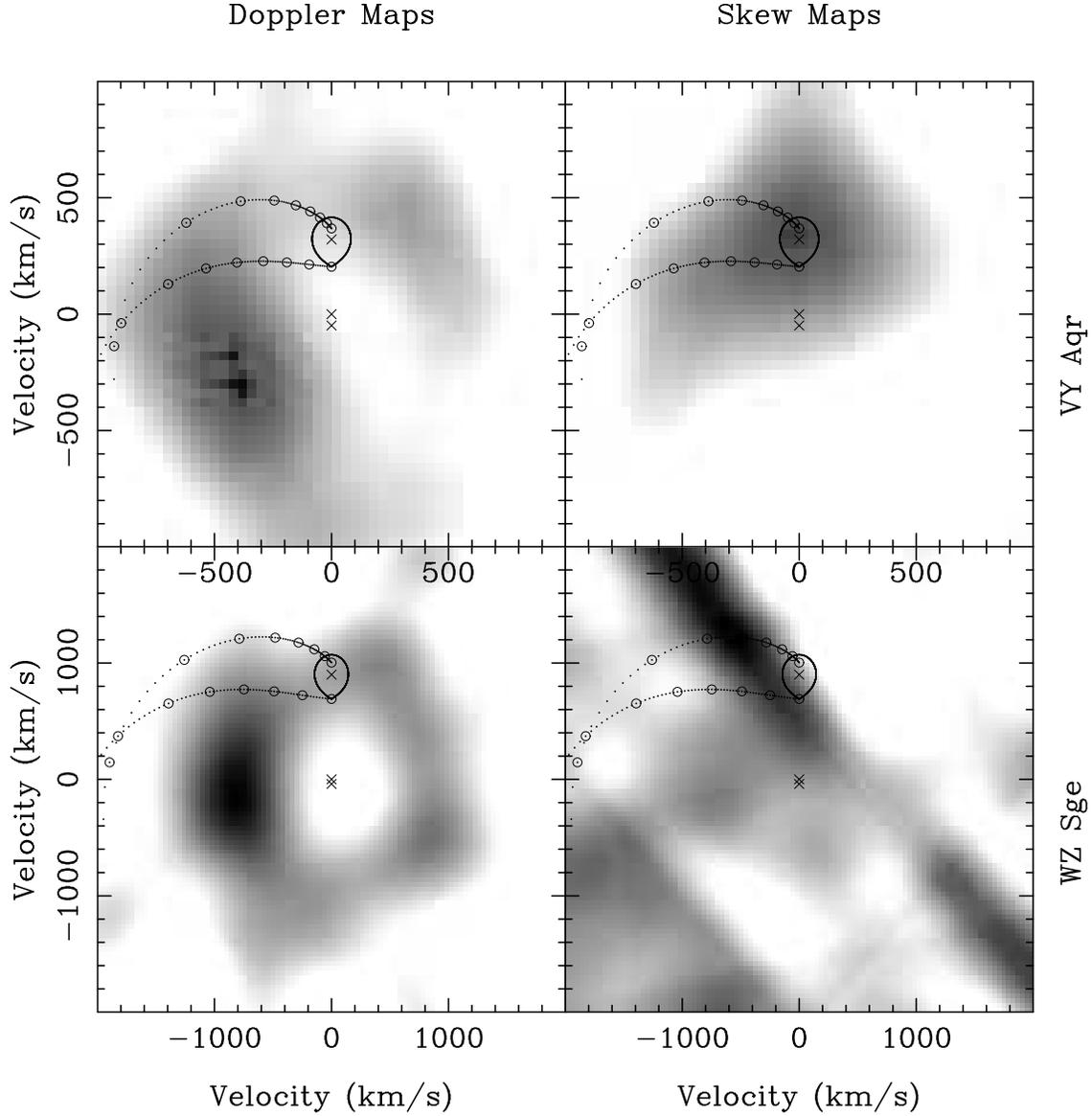}}
\caption[]{(Top) Left-hand panel: The Doppler map of the Paschen-$\beta$ line 
in VY~Aqr. Right-hand panel: The skew map of VY~Aqr, made with the M-dwarf 
Gl866AB~(M5V). 
The value of $K_R$ is given by $K_R^2=K_X^2+K_Y^2$ where $K_X$ and $K_Y$ are
found from the skew map by the position of the peak. 
Both maps have had a phase correction of --$0.28$ made to the ephemeris
of \protect\scite{thorstensen97}. The predicted position of the secondary star,
the path of the gas stream (lower curve) and the Keplerian velocity at the
gas stream (upper curve) are marked on both maps, using the value of $q$ 
derived in the text. The three crosses on the maps are, from top to bottom, the centre of mass of the secondary star, the system (at zero velocity) and the 
white dwarf. The circles plotted on the gas stream represent (from left to 
right) increasing distance from the white dwarf (in steps of one-tenth of the
distance to the inner Lagrangian point, $L_1$). 
\label{fig:vyaqrmaps}
(Bottom) Left-hand panel: The Doppler map of the Paschen-$\beta$
line in WZ~Sge. Right-hand panel: The skew map of WZ~Sge, made with the M-dwarf
Gl866AB~(M5V). The predicted position of the secondary star, the path of the 
gas stream and the Keplerian velocity at the gas stream are marked on both
maps, along with
the centre of mass of the system components, in an identical fashion to that
above, using the system parameters of \protect\scite{spruit98}.}
\label{fig:wzsgemaps} 
\end{figure*}

Skew maps were produced of both WZ~Sge and VY~Aqr. Skew mapping is a 
tomographic technique used to extract a value for the radial velocity of the 
secondary star, $K_R$.
It is particularly useful when the spectral features are too weak for 
conventional cross-correlation techniques to work. It is described in 
detail in \scite{smith93b} and has been successfully employed by \scite{sad98b}.
We compared regions of the spectra unaffected by emission lines with the 
spectra of our M5 and M7 dwarf stars.
The first step was to remove the continuum from the spectrum of both the CV
and the template stars.
This is done by dividing by a first order polynomial fit, 
and then subtracting a higher order fit to the continuum. This ensures that
line strength is preserved along the spectrum.
Each of the time-resolved CV spectra were then cross-correlated with the 
template star spectra, giving a time-series of cross-correlation functions 
(CCFs) for each template.
The skew maps were then generated by back-projecting the CCFs, in an identical
fashion to time-resolved spectra in standard Doppler tomography 
\cite{marsh88b}. If there is a detectable secondary star we would expect a
peak at (0, $K_R$) in the skew map. 

Figure \ref{fig:vyaqrmaps} shows the skew map of
VY~Aqr, made with Gl866AB and a standard Doppler map of the Paschen-$\beta$
line in VY~Aqr. 
When we first performed the skew mapping of VY~Aqr, a strong peak appeared in 
the  lower right-hand quadrant. Because we only cross-correlate with absorption
features in the M-dwarf template, it is unlikely that the peak could be from 
anything
but the secondary star, and so we believe that the unusual position of the
peak is due to a discrepancy
between the ephemeris of \scite{thorstensen97} and the actual ephemeris at
the time of our observations. \scite{thorstensen97} give their error in the 
period as $\pm 0.00004$ days, allowing for a discrepancy of up to seven orbits
between the calculated and actual phases at the time of observation.
Hence, both maps of VY~Aqr have had a phase correction of --$0.28$ applied to 
the ephemeris of \scite{thorstensen97}, in order to move the peak in the skew
map to the expected position on the K$_X$=0 axis.

\begin{figure}
\psfig{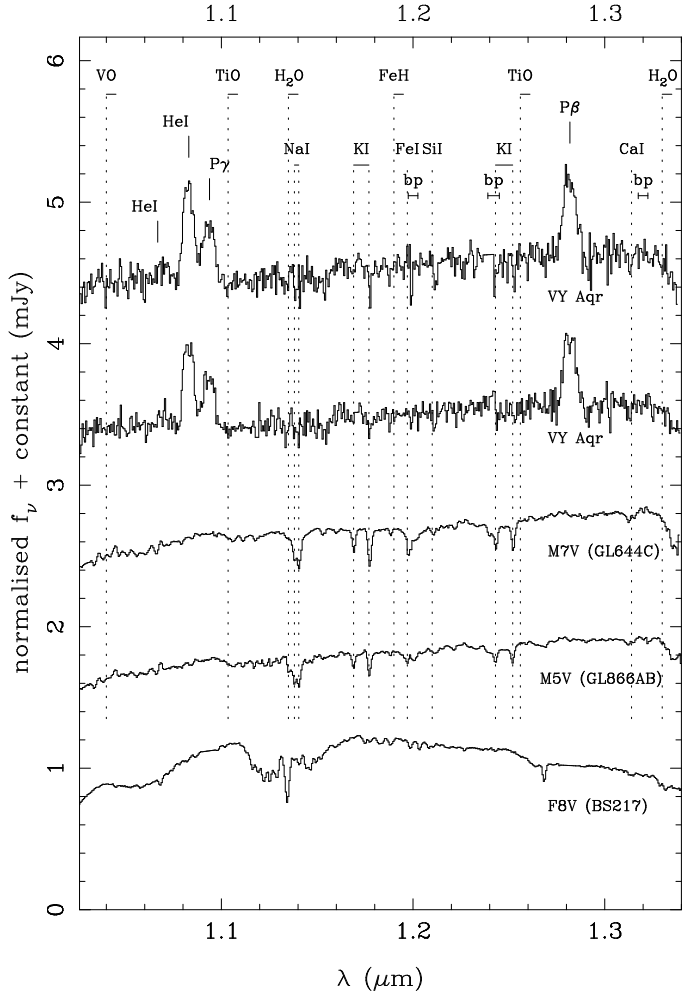}
\caption[]{Average J-band spectra of the short period DNe VY~Aqr and the
M dwarfs Gl644C and Gl866AB. The upper spectrum of VY Aqr has been
corrected for the radial velocity of the secondary star and represents an
average spectrum in the rest frame of the secondary star. The spectra have
been normalised by dividing by the flux at 1.3 $\mu$m and then offset by
adding a multiple of 0.9 to each spectrum. Also shown is the spectrum of an
F8V star, normalised by dividing by the flux at 1.3 $\mu$m, which indicates
the location of telluric features.}
\label{fig:velshift}
\end{figure}

The skew map of VY~Aqr shows a significant peak, yielding a value of 
$K_R=320\pm70$ kms$^{-1}$. As a check, the time resolved spectra of
VY~Aqr were shifted to correct for the radial velocity of the secondary
star and then averaged. The results are shown in figure~\ref{fig:velshift}.
Secondary star absorption features are clearly visible in the corrected 
spectrum, which also shows a much stronger and sharper water band. We 
conclude that our value of $K_R$ is appropriate, and not an artifact of the
skew mapping process.
\scite{thorstensen97} finds $K_W=49\pm4$ kms$^{-1}$ from 
H$\alpha$, giving a value for the mass ratio of the system of $q=0.15\pm0.04$.
This value of $q$ is consistent  with the precessing-disc model for SU UMa 
stars \cite{whitehurst88}, to which VY~Aqr belongs.
We are unable to derive component masses, as the inclination of the system is
uncertain \cite{augusteijn94}. Furthermore, it should be noted that $K_W$, 
when derived from emission lines,
does not always reflect the true motion of the white dwarf.

The Doppler map of VY~Aqr shows a faint ring corresponding to disc emission, 
and a peak we associate with the bright spot at (--500, --300) kms$^{-1}$. 
Assuming our phase offset is correct, we find that the bright spot in VY~Aqr 
is located downstream from the stream impact region.
This is consistent with the optical and infra-red doppler maps of WZ~Sge (see 
below) and with infra-red eclipse maps of IP~Peg, which show that the infrared
continuum bright spot can also appear downstream of it's optical counterpart 
\cite{froning99}. 
Work is in progress to determine if this is a common feature in infra-red 
accretion disc maps \cite{littlefair99}.
This consistency with other infrared  maps, as well as consistency with the 
radial velocity curves and light curves presented later in this paper gives us 
confidence that our phase correction is appropriate.

Figure \ref{fig:wzsgemaps} shows a standard doppler map of the Paschen-$\beta$
line in WZ~Sge, along with a skew map of WZ~Sge, made with Gl866AB. The doppler
map shows faint accretion disc structure and is dominated by emission from
the bright spot at (--800, --200) kms$^{-1}$. 
\scite{spruit98} found that the optical bright spot 
was located downstream from the stream impact region. Using their ephemeris
we find that the
infra-red bright spot is not aligned with it's optical counterpart but is
located further downstream from the stream impact region, consistent with
our results for VY~Aqr. 
The skew map of WZ~Sge shows no peak from the secondary star but a bright 
streak corresponding to a large value of the cross-correlation function at
$\phi \approx 0.9$. It is possible that this is due to correlation with
secondary star features, especially as it occurs near eclipse, 
although the absence of a strong correlation at other
phases, and the fact that it is not exactly at $\phi=0$ makes this unlikely.

\subsection{Light Curves}
\label{subsec:light curves}
\begin{figure*}
\centerline{\psfig{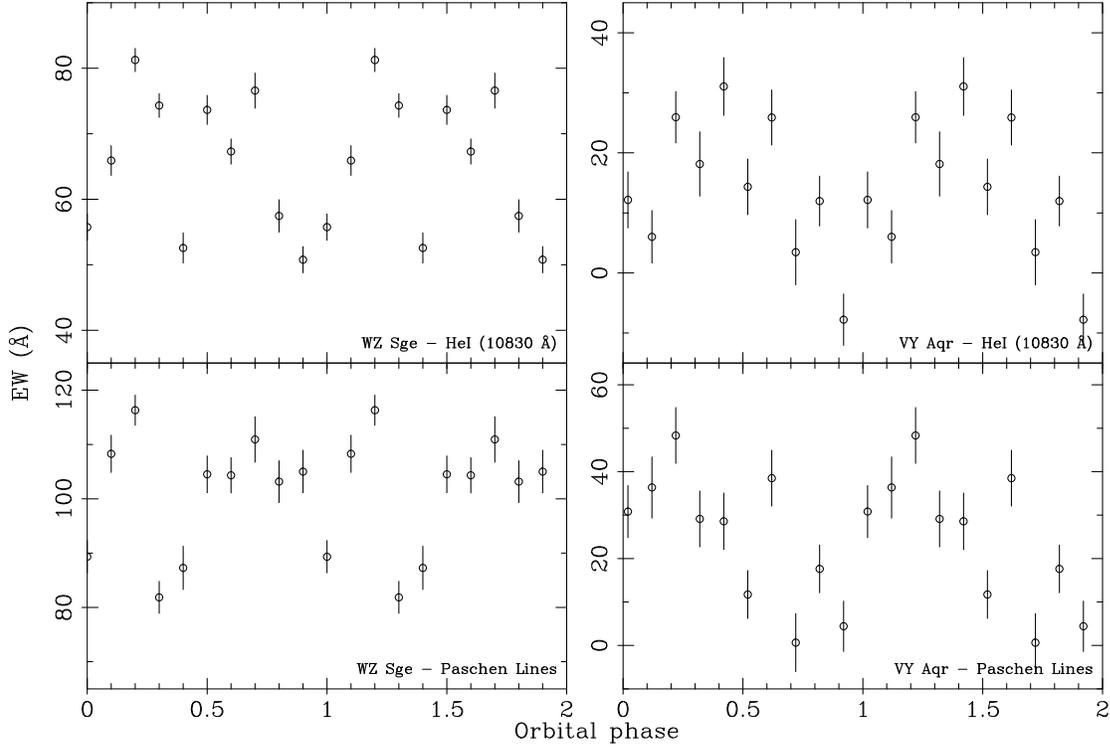}}
\caption[]{Emission line equivalent width (EW) light curves of WZ~Sge and 
VY~Aqr. EWs were calculated from the He\protect{\small I} (10830 \AA) 
line and from the sum of the Paschen lines (P$\gamma$ $+$ P$\beta$).
The He{\small I} line was separated from the Paschen-$\gamma$ line by 
measuring the EW up to a point at the middle of the blend. Hence the
He{\small I}  light curve is contaminated with a small amount of light from
Paschen-$\gamma$ and vice-versa.
The light curves have been folded over two complete 
orbits for clarity. Phases for WZ~Sge are from the ephemeris of 
\protect\scite{spruit98}, phases for VY~Aqr are shown using the 
ephemeris of \protect\scite{thorstensen97}, with a correction of --0.28 
applied.}
\label{fig:light}
\end{figure*}

Figure \ref{fig:light} shows equivalent width (EW) light curves for WZ~Sge and 
VY~Aqr. The light curves have been folded over two complete orbits for clarity.
They show a large amount of scatter, due to noise in the spectra, and hence
should be interpreted with caution.
Both the He{\small I} (10830 \AA) and the Paschen-lines light curve in WZ~Sge
show an eclipse at $\phi \approx 0$. 

The VY~Aqr light curves show a systematic variation in brightness, with a 
minimum brightness occurring at $\phi \approx 0.9$ in the He{\small I} lines
and at $\phi \approx 0.75$ in the Paschen lines, where we have applied
the correction of --0.28 to the ephemeris of \scite{thorstensen97}.
Such a variation could be produced either by obscuration of the bright spot
by an optically thick disc, or by self-obscuration of an optically thick
bright spot whose outer edge is brighter than the inner. 
Doppler maps of the He{\small I} emission in VY~Aqr show emission from a 
bright spot which is not aligned with the Pa$\beta$ emission. This explains 
the lag between the He{\small I} light curve and the Paschen-line light curve.
The presence of any variation at the orbital period in the light curves is 
surprising, as \scite{patterson93} found no variations on the orbital period 
in the optical continuum light curve of VY~Aqr. 
Further infrared photometry seems desirable. 

\subsection{Radial velocity curves}
\label{subsec:rvcurves}

\begin{figure*}
\centerline{\psfig{figure=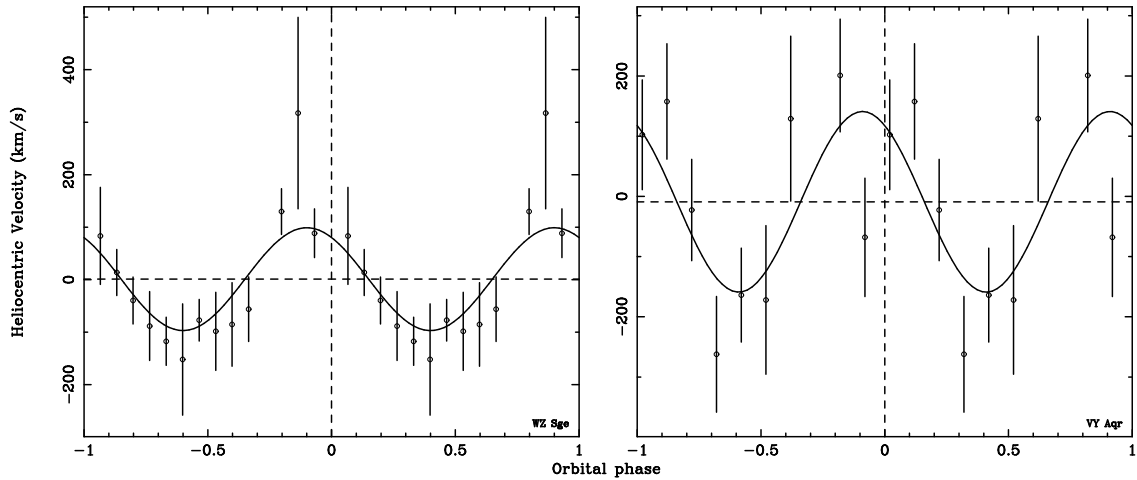,width=15.0cm,angle=0}}
\caption[]{Radial velocity curves of Paschen-$\beta$ for WZ~Sge and VY~Aqr,
measured using a double-Gaussian fit with Gaussian separations of 
1600 kms$^{-1}$ (VY~Aqr) and 2000 kms$^{-1}$ (WZ~Sge). Points at primary 
eclipse are omitted from the plot.
The curves are folded over two complete orbits for clarity.
Also shown are the fits to the curves (solid line) and the
systemic velocities (horizontal dashed line). Phases for WZ~Sge are from the 
ephemeris of \protect\scite{spruit98}, phases for VY~Aqr are shown using the 
ephemeris of \protect\scite{thorstensen97}, with a correction of --0.28 
applied.}
\label{fig:rvcurves}
\end{figure*}

\begin{figure*}
\centerline{\psfig{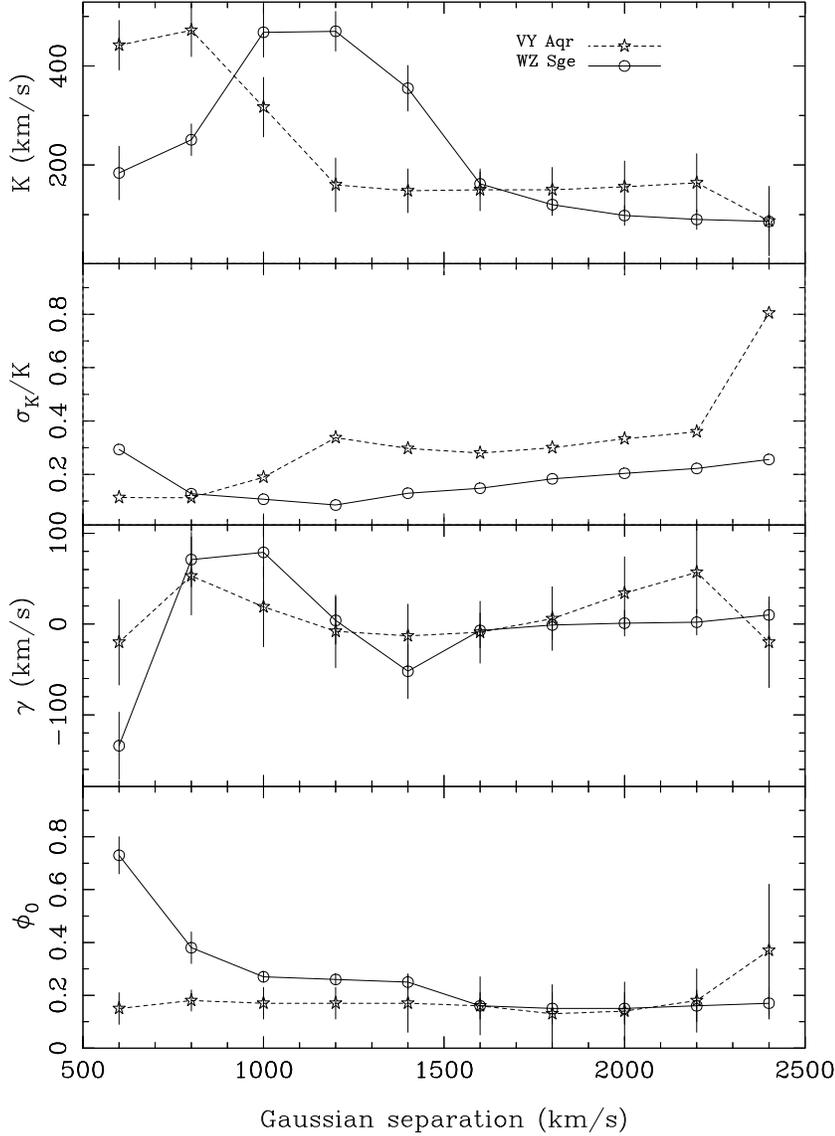}}
\caption[]{Diagnostic diagrams for WZ~Sge and VY~Aqr based on the 
double-Gaussian fits to Paschen-$\beta$. Phases for WZ~Sge are from the 
ephemeris of \protect\scite{spruit98}; phases for VY~Aqr are shown using the 
ephemeris of \protect\scite{thorstensen97}, with a correction of --0.28 
applied.}
\label{fig:diag}
\end{figure*}
 
The time resolved spectra were continuum subtracted and binned onto a constant
velocity interval scale. In order to measure the radial velocities, we applied
the double-Gaussian method of \scite{schneider80} to the Paschen-$\beta$ line.
This technique is sensitive mainly to the motion of the line wings and should 
therefore reflect the motion of the white dwarf with the highest reliability. 
The Gaussians were of width 500 kms$^{-1}$ (FWHM) and we varied their 
separation from 600 to 2400 kms$^{-1}$. We then fitted
\begin{equation}
V = \gamma -K\sin(\phi-\phi_0)
\end{equation}
to each set of measurements, omitting the points at primary eclipse. Examples
of the radial velocity curves obtained for VY~Aqr and WZ~Sge are shown 
in fig. \ref{fig:rvcurves}. 

The results of the radial velocity analysis are displayed in the form of a 
diagnostic diagram in figure \ref{fig:diag}. By plotting $K$, its fractional
error $\sigma_K/K$, $\gamma$ and $\phi_0$ as functions of the Gaussian 
separation it is possible to select the value of $K$ which most closely matches
$K_W$ \cite{shafter86}. If emission is disc dominated, one would expect $K$
to asymptotically approach $K_W$ when the Gaussian separation becomes 
sufficiently large. Furthermore, one would expect $\phi_0$ to approach 0.

For WZ~Sge it can be seen that the phase shift $\phi_0$ does indeed fall
towards 0 with increasing Gaussian separation. The value of $K$ steadily
drops, reaching its lowest value at around $K = 85 \pm 20$ kms$^{-1}$. 
At this point the phase shift is $0.17 \pm 0.06$ and not zero, as is typical 
for this sort of measurement in CVs (see e.g.\ section 2.7.6 in 
\pcite{warner95a} for a discussion). 
This phase shift introduces doubt about the validity of adopting this figure 
for $K_W$. Several authors have attempted to measure $K_W$ in WZ~Sge
with optical values ranging from 40-70 kms$^{-1}$ for H$\alpha$ and 60-80
kms$^{-1}$ for H$\beta$ -- see \scite{skidmore99}. \scite{spruit98}, using
a light centres method which attempts to correct for contamination by light
from the hotspot, find a value for H$\alpha$ of $40 \pm 10$ kms$^{-1}$.
K-band infrared studies of WZ~Sge yield a value for $K$, from the Br$\gamma$
line, of $108 \pm 15$ kms$^{-1}$ \cite{skidmore99}. 
Our value is compatible with the higher optical values.

For VY~Aqr the phase shift does not fall to zero, instead remaining roughly 
constant at $\phi_0 \approx 0.16$, again typical for this type of measurement.
The value of $K$ remains fairly constant at $K = 150 \pm 50$ kms$^{-1}$.
This value is inconsistent with the value obtained by 
\scite{thorstensen97} of $K_W = 49 \pm 4$ kms$^{-1}$ for H$\alpha$.

\begin{figure*}
\centerline{\psfig{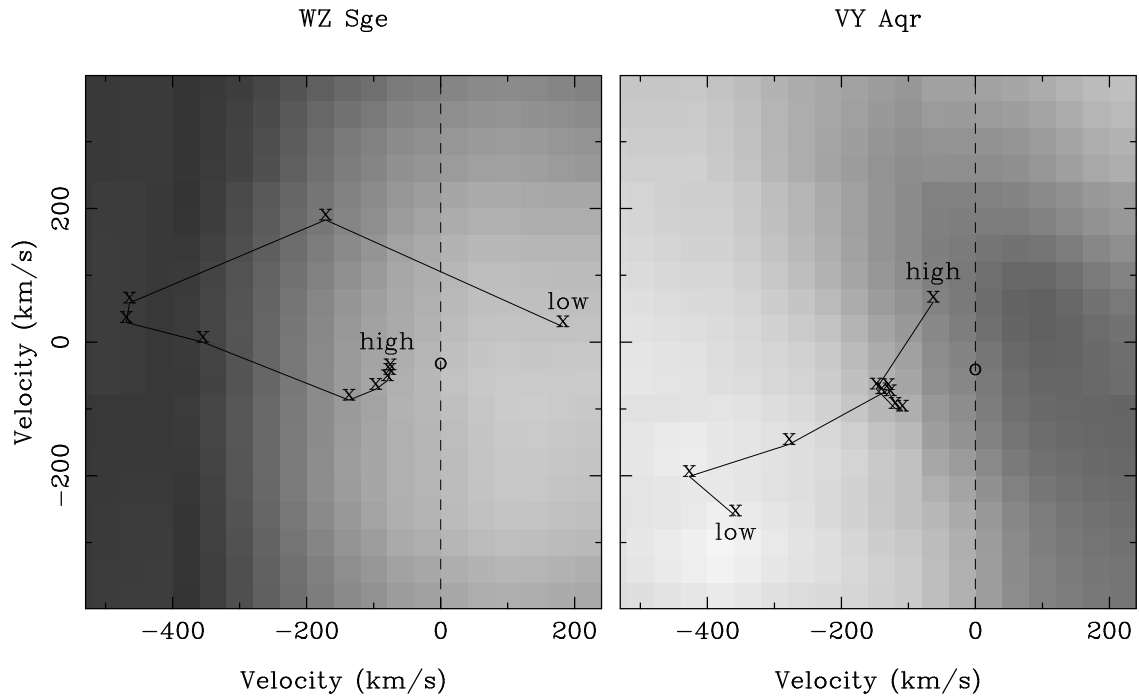}}
\caption[]{Light centres of the Paschen-$\beta$ emission in WZ~Sge and VY~Aqr,
superimposed on the Paschen-$\beta$ Doppler maps. 
For clarity, strong emission is represented as dark areas for WZ~Sge, and light
areas for VY~Aqr. Points are plotted for double Gaussian fits with separations 
ranging from 600 kms$^{-1}$ (marked ``low'') to 2400 kms$^{-1}$ 
(marked ``high''). The dashed line represents $K_X=0$, and is where the 
white dwarf is expected to lie. The centre of mass of the white dwarf 
(marked ``o'') is plotted using the values for $K_W$ of \scite{spruit98} 
for WZ~Sge and \scite{thorstensen97} for VY~Aqr. 
Phases for WZ~Sge are from the ephemeris of \protect\scite{spruit98}; phases 
for VY~Aqr use the ephemeris of \protect\scite{thorstensen97}, 
with a correction of --0.28 applied.}
\label{fig:lightcent}
\end{figure*}

It is not easy to read reliable values of $K_W$ from our diagnostic diagram.
Thus, in an attempt to obtain reliable values for $K_W$ for both VY~Aqr and 
WZ~Sge we employed a modified version of the light centres method, described 
by \scite{marsh88a}. In the co-rotating co-ordinate frame, the white dwarf has
velocity (0,-$K_W$), and any symmetric emission (e.g.\ from the disc) would
be centred at that point. By plotting $K_X = -K \sin \phi_0$ against 
$K_Y = -K \cos \phi_0$ for the different radial velocity fits, we find that
the points move closer to the $K_Y$ axis with increasing Gaussian separation.
This is because at large separations the fits are dominated by the line
wings, which should be formed in the symmetric, inner disc. An extrapolation
of the points to the $K_Y$ axis should give a measurement of $K_W$.
Such a light centres diagram is presented in figure \ref{fig:lightcent}.

The light centres in both plots seem to cluster at the edges of the bright 
spots as the gaussian seperation increases. Hence we are unable to
extrapolate to $K_X = 0$ and therefore unable to estimate $K_W$ for
WZ~Sge and VY~Aqr.

\section{Discussion}
\label{sec:discussion}

\subsection{Detection of secondary star features}
\label{subsec:detection}

We consider here the reality of the secondary star detection in VY~Aqr and 
EF~Peg. The secondary star detection is based on three points of evidence; the
continuum slope, the presence of the water band at 1.33 $\mu$m and the success
of the skew mapping technique. We consider these points in turn.

\begin{enumerate}
\item{VY~Aqr has a red continuum. The continuum shows a change in slope around
1.1 $\mu$m, a feature shared by the M-dwarf spectra.
In fact, the continuum shape in VY~Aqr strongly resembles that of the M-dwarfs.
As the accretion disc and white dwarf
are expected to have a blue continuum, it is reasonable to ascribe the red 
continuum seen in VY~Aqr to light from the secondary star. This does
not necessarily imply that the secondary dominates the IR light, as the
amount of secondary star light needed to create an overall red continuum
is strongly dependant upon the slope of the secondary star spectrum. 
The continuum slope in VY~Aqr appears to show no variation with orbital phase, 
always being within one standard error of the mean.}

\item{The water band at 1.33 $\mu$m lies in a region affected by telluric
features (see fig. \ref{fig:cvav}), so it is possible that incorrect telluric 
correction could affect this feature. In order to try and estimate the 
extent of this effect, spectra of VY~Aqr
were calibrated using each of the F-stars observed. It was seen that 
there was excess absorption at 1.33 $\mu$m, even when the stronger telluric 
features around 1.14 $\mu$m were over corrected. Therefore, incorrect 
telluric correction cannot account for the observed feature.
The presence of the unexplained variations in strength of the water feature 
with orbital phase (as shown in figure \ref{fig:tr}) is disturbing, however. 
Following the method outlined above, we find that incorrect telluric
correction cannot account for this variation. Also, by repeating the reduction
process with a range of parameters and methods, we determine that
systematic errors in the reduction process do not account for the variation
either.
The feature is only present in the time resolved spectra at a significance
of one standard error, so we might attribute the variation in the band
to random noise.
However, the detection of this feature cannot be regarded as certain. It is
noted here that the percentage contributions of the secondary to the J-band
light in VY~Aqr and EF~Peg are dependant on this feature, and are hence also
uncertain.}

\item{The skew mapping technique correlates with a secondary star template.
Strong cross-correlations are obtained with this method. When the
time-resolved spectra are averaged in the rest frame of the secondary star,
absorption features are stronger and sharper.
Further, the skew
mapping suggests a correction to the ephemeris for VY~Aqr, which is consistent
with the radial velocity curves and light curves. Thus we are confident that
the skew mapping is correlating with real secondary star features.}
\end{enumerate}

None of these items, considered alone, provide concrete signs of the secondary
star. However, in combination they provide a strong body of evidence.
Hence we are confident that the secondary star has been detected in VY~Aqr.
The detection in EF~Peg is based only on the continuum shape and the water 
feature at 1.33 $\mu$m. Hence the detection of the secondary in this case is 
less reliable than that in VY~Aqr.

\section{Conclusions}
\label{sec:conclusions}

\begin{enumerate}

\item{The relative ease of observation in the J-band, and strong absorption 
features of late-type dwarfs in this spectral region make J-band studies highly
desirable for CVs with undetected secondary stars.}

\item{The secondary star has been detected in VY~Aqr and EF~Peg,
although the detection was too weak, and insufficient field-dwarf templates 
were observed, to make an estimate of the spectral type.}

\item{The contribution of the secondary star to the total J-band
flux was found to be $65\%$ for EF~Peg and $70\%$ for VY~Aqr. 
Upper limits of $10\%$ and $20\%$ were found for the contribution
of the secondary to the total J-band flux for WZ~Sge and PX~And respectively.} 

\item{A value of $K_R = 320\pm70$ kms$^{-1}$ was found for VY~Aqr and 
the mass ratio was calculated as $q=0.15\pm0.04$, assuming $K_W$, as derived
by \scite{thorstensen97} is correct.}

\item{The spectral type of WZ~Sge was constrained to be later
than M7.5V.}

\end{enumerate}

\section*{\sc Acknowledgements}

We are indebted to Mark Huber for supplying J-band magnitudes for this paper, 
and Tom Marsh for use of his software package, {\sc molly}.
We wish to thank the referee, Prof. Tim Naylor for his constructive
comments.
SPL is supported by a PPARC studentship.
UKIRT is operated by the Joint Astronomy Centre on 
behalf of the Particle Physics and Astronomy Research 
Council. The authors acknowledge the data analysis facilities at Sheffield
provided by the Starlink Project which is run by CCLRC on
behalf of PPARC
\bibliographystyle{mnras}
\bibliography{refs}

\begin{table*}
\caption[]{Journal of observations. Each spectrum consists of 240 s total
exposure time.
 Gaps between objects were used for arcs, 
flats and F-star spectra. An object is listed twice where such a gap broke 
continuous observation. The long gap between the observation of Gl866AB
and WZ~Sge is due to cloud. Observations of VY~Aqr between 07:20 and 08:54
contained no useful information, due to very poor seeing.
Spectral types are those given by \protect\scite{kirkpatricketal94}. 
Orbital phase is shown for WZ~Sge using the ephemeris of \scite{spruit98}, 
PX~And using the ephemeris of \scite{still95b}
and VY~Aqr using the ephemeris of \scite{thorstensen97}.
The orbital period for EF~Peg is estimated from the superhump period
 \protect\cite{howell93}.}
\begin{center}
{\bf
\begin{tabular}{@{\extracolsep{-2.15mm}}lcccccccccc}
& & & & & & & & &\\
\multicolumn{1}{l}{Object} &
\multicolumn{1}{c}{$\lambda_{\rm central}$} &
\multicolumn{1}{c}{Date} &
\multicolumn{1}{c}{UT} &
\multicolumn{1}{c}{UT} &
\multicolumn{1}{c}{No. of} &
\multicolumn{1}{c}{Period} &
\multicolumn{1}{c}{Phase} &
\multicolumn{1}{c}{Phase} &
\multicolumn{1}{c}{Spectral} &
\multicolumn{1}{c}{J-band}\\
 & \multicolumn{1}{c}{($\mu$m)} & & 
\multicolumn{1}{c}{start} &
\multicolumn{1}{c}{end} &
\multicolumn{1}{c}{Spectra} &
\multicolumn{1}{c}{(days)} &
\multicolumn{1}{c}{start} &
\multicolumn{1}{c}{end} & 
\multicolumn{1}{c}{Type} & 
\multicolumn{1}{c}{mag.}\\
& & & & & & & & &\\
WZ Sge        & 1.175 & 08/08/98 & 06:40 & 08:33 & 24 & 0.0567 & 0.24 & 1.43 &               & 15.3 \\ 
WZ Sge        & 1.175 & 08/08/98 & 08:47 & 10:36 & 24 & 0.0567 & 1.80 & 2.97 &               & 15.3 \\ 
Gl866AB       & 1.175 & 08/08/98 & 10:56 & 11:27 &  4 & & & & M5V & \\
WZ Sge        & 1.175 & 08/08/98 & 12:40 & 13:50 & 14 & 0.0567 & 4.66 & 5.19 &               & 15.3 \\
PX And        & 1.175 & 08/08/98 & 14:09 & 15:04 & 12 & 0.1464 & 0.61 & 0.82 &               & 14.2 \\
Gl644C        & 1.175 & 09/08/98 & 06:20 & 06:45 &  4 & & & & M7V & \\
VY Aqr        & 1.175 & 09/08/98 & 07:20 & 08:54 & 20 & 0.0635 & 0.81 & 1.19 &               & 14.2 \\
VY Aqr        & 1.175 & 09/08/98 & 09:22 & 10:53 & 20 & 0.0635 & 1.50 & 2.36 &               & 14.2 \\
EF Peg        & 1.175 & 09/08/98 & 11:36 & 13:09 & 20 & 0.0837 & & & &\\
EF Peg        & 1.175 & 09/08/98 & 13:33 & 14:46 & 16 & 0.0837 & & & &\\
& & & & & & & &\\
\end{tabular}
}
\end{center}
\label{tab:journal}
\end{table*}

\clearpage
\newpage

\begin{table*}
\caption[]{Wavelengths, equivalent widths and velocity widths of the most
prominent lines in the J-band spectra of the novalike variable PX~And,
the DNe WZ~Sge, EF~Peg, VY~Aqr and the M-dwarfs Gl644C and Gl866AB. 
The line identifications were made using the lists and spectra of 
\pcite{kirkpatricketal93} (marked $^a$), 
\pcite{joyce98} (marked $^b$),
\pcite{jones94} (marked $^c$), \pcite{jones96} (marked $^d$),
\pcite{tinney93} (marked $^e$), \pcite{lang86} (marked $^f$) and the UKIRT 
online documentation (marked $^g$). Equivalent widths for the water band
around 1.34 $\mu$m were measured in the range 1.32--1.34 $\mu$m. 
The vertical bars following lines of similar wavelength indicate that the 
measurements apply to the entire blend. Wavelengths for VO, TiO, H$_2$O and FeH
refer to the band heads, or the edge of the band, if headless. 
The two-letter codes indicate that the line is either not 
present (np) or that the line is present but is not measurable (nm).} 
\newcommand{\vd}{$\pm$}
{\hspace*{-0.5cm}
\begin{tabular}{@{\extracolsep{-2.15mm}}lccccccccccccc}
& & & & & & & & & & & & &\\
& & \multicolumn{2}{c}{\bf WZ Sge} & \multicolumn{2}{c}{\bf EF Peg} & 
\multicolumn{2}{c}{\bf VY Aqr} & \multicolumn{2}{c}{\bf PX And} &
\multicolumn{2}{c}{\bf Gl866AB} &
\multicolumn{2}{c}{\bf Gl644C} \\
{\bf Line} & \multicolumn{1}{c}{$\lambda$} & \multicolumn{1}{c}{\bf EW} & 
\multicolumn{1}{c}{\bf FWHM} 
& \multicolumn{1}{c}{\bf EW} & \multicolumn{1}{c}{\bf FWHM} & 
\multicolumn{1}{c}{\bf EW} & \multicolumn{1}{c}{\bf FWHM} & 
\multicolumn{1}{c}{\bf EW} & \multicolumn{1}{c}{\bf FWHM} &
\multicolumn{1}{c}{\bf EW} & \multicolumn{1}{c}{\bf FWHM} &
\multicolumn{1}{c}{\bf EW} & \multicolumn{1}{c}{\bf FWHM} \\
& \multicolumn{1}{c}{($\mu$m)} & \multicolumn{1}{c}{\bf \AA} & \multicolumn{1}{c}
{\bf km\,s$^{-1}$} 
& \multicolumn{1}{c}{\bf \AA} & \multicolumn{1}{c}{\bf km\,s$^{-1}$} & 
\multicolumn{1}{c}{\bf \AA} & \multicolumn{1}{c}{\bf km\,s$^{-1}$} & 
\multicolumn{1}{c}{\bf \AA} & \multicolumn{1}{c}{\bf km\,s$^{-1}$} & 
\multicolumn{1}{c}{\bf \AA} & \multicolumn{1}{c}{\bf km\,s$^{-1}$} & 
\multicolumn{1}{c}{\bf \AA} & \multicolumn{1}{c}{\bf km\,s$^{-1}$} \\
& & & & & & & & & & & \\
{\bf HeI}            & {\bf 1.0830$^f$}       & 67\vd1   & 2300\vd100 & 28\vd3  & 1500\vd300 & 53\vd2   & 1500\vd100 & 46\vd1 & 890\vd30 & np & np & np & np    \\
{\bf Pa-$\gamma$}    & {\bf 1.0938$^f$}       & 34\vd1   & 2100\vd100 & 15\vd3  & 1700\vd300 & 30\vd2   & 1500\vd100 & 51\vd1 & 1200\vd30 & np & np & np & np   \\
{\bf Pa-$\beta$}     & {\bf 1.2818$^g$}       & 69\vd1   & 2400\vd100 & 30\vd3  & 2100\vd300 & 40\vd3   & 1600\vd100 & 58\vd2 & 1200\vd30 & np & np &np & np    \\
{\bf HeI}            & {\bf 1.0668$^g$}       & 11\vd1   & nm         & nm      & nm         & np       & np         & np     & np        & np & np &np & np    \\
{\bf VO}             & {\bf 1.0400$^b$}       & np       & np         & np      & np         & np       & np         & np      & np         & nm      & nm     & nm         & nm    \\
{\bf TiO}            & {\bf 1.1035$^b$}       & np       & np         & np      & np         & np       & np         & np      & np         & nm      & nm     & nm         &nm     \\ 
{\bf H$_{2}$0}       & {\bf 1.1350$^b$\vline} &          &            &         &            &          &            &         &            &         &        &            &       \\
{\bf NaI}            & {\bf 1.1381$^c$\vline} & np       & np         & np      & np         & np       & np         & np      & np         & -14\vd1 & nm     & -10\vd1    & nm    \\
{\bf NaI}            & {\bf 1.1404$^c$\vline} &          &            &         &            &          &            &         &            &         &        &            &       \\
{\bf KI}             & {\bf 1.1690$^c$ \vline} & np       & np         & np      & np         & np       & np         & np      & np         & -6\vd1  & nm     & -12\vd1    & nm    \\
{\bf KI}             & {\bf 1.1773$^c$ \vline} &          &            &         &            &          &            &         &            &         &        &            &       \\
{\bf FeH}            & {\bf 1.1900$^a$\vline} & np       & np         & np      & np         & np       & np         & np      & np         & -7\vd1  & nm      & -15\vd1    & nm    \\
{\bf FeI}            & {\bf 1.1970$^d$\vline} &          &            &         &            &          &            &         &            &         &        &            &       \\
{\bf SiI}            & {\bf 1.2100$^e$}       & np       & np         & np      & np         & np       & np         & np      & np         & -2\vd1  & nm     & nm         & nm    \\
{\bf KI}             & {\bf 1.2432$^c$\vline} & np       & np         & np      & np         & np       & np         & np      & np         & -10\vd1 & nm     & -21\vd1    & nm    \\
{\bf KI}             & {\bf 1.2522$^c$\vline} &          &            &         &            &          &            &         &            &         &        &            &       \\
{\bf TiO}            & {\bf 1.2560$^b$}       & np       & np         & np      & np         & np       & np         & np      & np         & -1\vd1  & nm      & nm         & nm   \\
{\bf CaI}            & {\bf 1.3140$^c$}       & np      & np          & np      & np         & np       & np         & np      & np         & -1\vd1  & nm      & -1\vd1     & nm   \\
{\bf H$_2$O}         & {\bf 1.3400$^c$}       & np      & np          & -14\vd4 & nm         & -17\vd4  & nm         & np      & np         & -10\vd1 & nm      & -14\vd1    & nm    \\
& & & & & & & & & & & \\
\end{tabular}
}
\label{tab:lines}
\end{table*}

\begin{table*}
\newcommand{\vd}{$\pm$}
\caption[]{The Contributions of the secondary star to the total J-band flux.}
{\bf
\begin{tabular}{lcccc}
 & & & &\\
\multicolumn{1}{l}{Object} &
\multicolumn{1}{c}{Secondary} &
\multicolumn{1}{c}{Error} &
\multicolumn{1}{c}{M-dwarf} &
\multicolumn{1}{c}{Mag. of}\\ 
& \multicolumn{1}{c}{Contribution} &
& \multicolumn{1}{c}{used} &
\multicolumn{1}{c}{Secondary} \\
 & & \\
WZ~Sge	&  $\leq$10\% & 5\% & M7 & $\geq$16.2\\
PX~And  &  $\leq$20\% & 5\% & M5 & $\geq$15.4\\
EF~Peg  &        65\% & 15\% & M7 & 16.7\vd0.5\\
VY~Aqr  &        70\% & 15\% & M5 & 15.7\vd0.5\\
& & & & \\
& & & & \\
\end{tabular}
}
\label{tab:contrib}
\end{table*}

\end{document}